\documentclass[10pt, conference, letterpaper]{IEEEtran}
\usepackage{amsmath}
\usepackage[utf8]{inputenc}
\usepackage{pdfpages}
\usepackage{cite}
\usepackage{amsmath}
\usepackage{amssymb}
\usepackage{mathtools}
\usepackage{amsmath}
\usepackage{amstext}
\usepackage{amssymb}
\usepackage{amsfonts}
\usepackage{bm}
\usepackage{algorithm}
\usepackage{algorithmic}

\usepackage{tabularx}
\usepackage{latexsym}
\usepackage{paralist}
\usepackage{float}
\usepackage{enumerate}

\usepackage{paralist}
\usepackage{amsopn}
\usepackage{mathrsfs}
\usepackage{mathtools}
\usepackage{amsthm}
\usepackage{amssymb}
\usepackage{amsfonts}
\usepackage{comment}
\usepackage{xspace} 
\usepackage{relsize}
\usepackage{float}
\usepackage{bbm}
\usepackage{tkz-euclide}
\usepackage{pgfplots}
\usepackage{enumitem}

\usepackage{graphicx}
\usepackage{paralist}
\usepackage{xspace} 
\usepackage{multicol}
\usepackage{subcaption}
\usepackage[font=footnotesize]{caption}

\usepackage[official]{eurosym}

\graphicspath{{/Users/jonair/Dropbox/doc/publish/2/graphics/used/}}

\newcommand{\Files}{\mathcal{F}}

\newcommand{\fsize}{b}

\newcommand{\BS}{\mathcal{M}}
\newcommand{\bs}{m}

\newcommand{\mem}{k}

\newcommand{\numUT}{N}
\newcommand{\R}{\mathbb{R}}

\newcommand{\Z}{\mathbb{Z}}
\newcommand{\noise}{\sigma^2}

\newcommand{\fa}{\hspace{5pt}\forall\hspace{2pt}}

\newcommand{\lcard}{\left\vert}
\newcommand{\rcard}{\right\vert}

\newcommand{\Regions}{\mathcal{S}}
\newcommand{\region}{s}

\newcommand{\price}{q}
\newcommand{\x}{x}
\newcommand{\y}{y}
\newcommand{\z}{z}
\newcommand{\lag}{\lambda}
\newcommand{\xvec}{\mathbf{\x}}
\newcommand{\yvec}{\mathbf{\y}}
\newcommand{\zvec}{\mathbf{\z}}
\newcommand{\lagvec}{\bm{\lag}}

\newcommand{\uapol}{\Pi}
\newcommand{\sur}{\textsc{sur}}

\newcommand{\pow}{p}
\newcommand{\chan}{h}
\newcommand{\XZ}{\mathcal{X}}
\newcommand{\Y}{\mathcal{Y}}
\newcommand{\Opt}{\textsc{Opt}\text{-}\ensuremath{\savings}\xspace}
\newcommand{\Closest}{\textsc{Closest}\xspace}

\DeclareMathOperator*{\argmax}{arg\,max}

\DeclareMathOperator{\coverBS}{\BS}
\DeclareMathOperator{\bsRegions}{\Regions}
\DeclareMathOperator{\util}{U}

\DeclareMathOperator{\savings}{h}

\DeclareMathOperator{\sinr}{\text{SINR}}

\newcommand{\benders}{v}
\newcommand{\bendersvec}{\mathbf{\benders}}
\newcommand{\vol}{v}
\newcommand{\w}{w}

\makeatletter
\let\amsmath@bigm\bigm

\renewcommand{\bigm}[1]{%
  \ifcsname fenced@\string#1\endcsname
    \expandafter\@firstoftwo
  \else
    \expandafter\@secondoftwo
  \fi
  {\expandafter\amsmath@bigm\csname fenced@\string#1\endcsname}%
  {\amsmath@bigm#1}%
}

\newcommand{\DeclareFence}[2]{\@namedef{fenced@\string#1}{#2}}
\makeatother

\DeclareFence{\mid}{|}
\IEEEoverridecommandlockouts

\begin{document}

\title{Optimal Cache Leasing from a Mobile Network Operator to a Content Provider}

%\author{\IEEEauthorblockN{Jonatan Krolikowski\IEEEauthorrefmark{1}, Anastasios Giovanidis\IEEEauthorrefmark{2} and Marco Di Renzo\IEEEauthorrefmark{1}} 
%\IEEEauthorblockA{\IEEEauthorrefmark{1}CNRS-L2S, CentraleSup{\'e}lec, Université Paris Sud, Universit{\'e} Paris-Saclay\\
%\IEEEauthorrefmark{2}CNRS-LIP6, Universit{\'e} Pierre et Marie Curie, Sorbonne Universit{\'e}s, Paris, France\\
%Email: jonatan.krolikowski@telecom-paristech.fr, anastasios.giovanidis@lip6.fr, marco.direnzo@lss.supelec.fr}}

%\author{Jonatan Krolikowski \footnote{CNRS-L2S, CentraleSup{\'e}lec, Université Paris Sud, Universit{\'e} Paris-Saclay} } %\affiliation{CentraleSup\'elec, Paris, France}
%\author{Anastasios Giovanidis}%\affiliation{CNRS -- T\'el\'ecom ParisTech, Paris, France}
%\author{} %\affiliation{CNRS -- CentraleSup\'elec, Paris, France}

\author{\IEEEauthorblockN{Jonatan Krolikowski\IEEEauthorrefmark{1}, Anastasios Giovanidis\IEEEauthorrefmark{2} and Marco Di Renzo\IEEEauthorrefmark{1}\thanks{This research is funded by Digiteo/Digicosme.}} 
\IEEEauthorblockA{\IEEEauthorrefmark{1}CNRS-L2S, CentraleSup{\'e}lec, Université Paris Sud, Universit{\'e} Paris-Saclay\\
\IEEEauthorrefmark{2}CNRS-LIP6, Universit{\'e} Pierre et Marie Curie, Sorbonne Universit{\'e}s, Paris, France\\
Email: jonatan.krolikowski@telecom-paristech.fr, anastasios.giovanidis@lip6.fr, marco.direnzo@lss.supelec.fr}}
%\IEEEauthorblockB{CNRS-LTCE, T{\'e}l{\'e}comParistech, Universit{\'e} Paris-Saclay\\
%Email: \IEEEauthorrefmark{2}anastasios.giovanidis@telecom-paristech.fr}

%\IEEEpeerreviewmaketitle
\maketitle

 %\pacs{}
\begin{abstract}
Caching popular content at the wireless edge is recently proposed as a means to reduce congestion at the backbone of cellular networks. The two main actors involved are Mobile Network Operators (MNOs) and Content Providers (CPs). In this work, we consider the following arrangement: an MNO pre-installs memory on its wireless equipment (e.g. Base Stations) and invites a unique CP to use them, with monetary cost. The CP will lease memory space and place its content; the MNO will associate network users to stations. For a given association policy, the MNO may help (or not) the CP to offload traffic, depending on whether the association takes into account content placement. 

We formulate an optimization problem from the CP perspective, which aims at maximizing traffic offloading with minimum leasing costs. This is a joint optimization problem that can include any association policy, and can also derive the optimal one. We present a general exact solution using Benders decomposition. It iteratively updates decisions of the two actors separately and converges to the global optimum. We illustrate the optimal CP leasing/placement strategy and hit probability gains under different association policies. Performance is maximised when the MNO association follows CP actions.
\end{abstract}

%\section{Completed work and context}

%This report presents the ongoing research on my thesis on the subject "Optimal Content Management and Dimensioning in Wireless Networks". The full scope of this research has not been published yet. Several possibilities to extend the presented work  are discussed in Section~\ref{section:conclusion}. One aspect of the general problem has been published in the WIOPT-CCDWN 2017 workshop \cite{Krolikowski2017}: Given a placement of content into caches of wireless nodes, how can the mobile network operator perform user association such that all caches are useful? This problem is a special  case of the user association problem ASSOC presented in Section~\ref{section:problem}, namely a formulation of ASSOC as a network utility maximization (NUM) problem which is further discussed in Section~\ref{section:cases}.

%In \cite{Krolikowski2017}, a distributed algorithm is developed which converges to the optimal solution of the NUM problem. The core of the algorithm is a novel sub-routine called \emph{bucket-filling}. When a fairness criterion is chosen as objective of the NUM,  cache-related traffic is diverted from potentially overloaded wireless nodes to less utilized ones.

\section{Introduction}
 
By 2020, wireless data traffic is estimated to reach roughly the 8-fold of its volume of 2015~\cite{Cisco2016}. Such increase is a challenge to mobile network operators (MNOs)  as well as to the content providers (CPs).
The MNOs and the CPs have different strategies to cope with these high demands: The MNOs, on the one hand, try to satisfy the increase in data demand  by densifying the network with new tiers
%Solutions such as  the  placement of additional base stations
%, extending the spectrum  
and by allowing cooperation among  stations. 
%are considered in order to increase the throughput of the wireless channel and  reduce delays. 
However, this increase of wireless traffic can pose new problems to the wireless backhaul  that are related to congestion. On the other hand, 
the  CPs are on the receiving end of the content requests. Large CPs such as Youtube or Netflix  store their data in huge data centers. For such a CP, a steep increase in data demand can be handled by  massive infrastructure investments, i.e.\  upgrade of the data center capacity as well as installation of higher capacity data links to the surrounding network.
 % Increasing backhaul capacity cannot be a solution to this problem because it is very costly. %\cite{Caire2013}. 
 
A recently studied alternative  is to equip wireless nodes with caches (see \cite{Caire2013,Blaszczyszyn2014,Bastug2015,Poularakis2016,Douros2017}). The main  purpose of cache memory installation is to ease  backhaul traffic and its processing at  the data centers by  handling content requests  from intermediate caches placed at the edge or inside  the core network.  %The content  is  brought closer to the user and the costly usage of backhaul bandwidth and data center capacity is reduced. 
This way, the  user Quality of Service (QoS) can be improved because cached content is downloaded with less delay from caches  closer to the user. 

Caching can be both of interest to the MNOs and the CPs. 
We consider a scenario in which an MNO has constructed and physically maintains caching infrastructure at Cached Base Stations (CBSs). The MNO makes three strategic decisions: 
\begin{compactitem}
\item[MNO-1)] How much memory to install at each CBS?
\item[MNO-2)] Which price to set for the leasing of one cache unit?
\item[MNO-3)] Which user association policy to pursue?
\end{compactitem}
The standard user association policy  routes each user to its closest station because of the strongest channel quality. More involved policies  allow load-balancing of users when these are located in the overlap of  coverage areas of two or more CBSs.
%In this work, we consider a policy which prioritizes the association of users to CBSs which have the requested content cached. If several such CBSs are available, the MNO balances the load of \emph{cache-related} traffic among the CBSs. Resource constraints of the CBSs are taken into account

A central assumption we make in this work is that the MNO puts the management of cache content into the hands of the CP. There are two reasons for this. The first is that the  CP usually transfers data to wireless users via secure connections, e.g.\ https \cite{Araldo2016}, so that the intermediate MNO cannot recognize requests and serve them from local caches.  %2)  Content placement policies may depend on  spatio-temporal  data about content popularity that the CP can gather and use.
 The second is that  content placement policies may depend on  spatio-temporal popularity  data that the CP can gather and use. All content related information is in the hands of the CP.

\emph{In what follows, we take the point of view of the CP}, which reacts to the MNO's cache installation, pricing scheme, and user association policy.  The CP has to take two types of decisions:
\begin{compactitem}
\item[CP-1)] How much cache space to lease from the MNO at each CBS? 
\item[CP-2)] Which content to place into the respective leased caches?
\end{compactitem}
 The decisions are  based on the estimated spatio-temporal popularity of the CP's content.
They are taken for a time period in which the system parameters are considered relatively static. For example, updates can take place in off-peak hours  \cite{Poularakis2014b}.
The aim of the CP is to find a cache leasing and content placement strategy which optimally weighs the total cache leasing costs against  the savings  from a reduction of traffic at the data centers: For every object stored in a cache, a price for  memory space and content delivery has to be paid. But,  at the same time, all users  that are associated to the cache and ask for the cached content  generate savings for the CP by relieving  data centers \cite{Roberts2013}. %Every expected user request that is served from a local cache  and that is thus not forwarded to a data center is beneficial  for the CP. 
To better understand the problem, consider two examples. In each case, the MNO installs unlimited cache memory at all CBSs and sets a price per memory unit. Then a user association policy is announced:

\emph{\Closest:} The MNO associates each wireless user to the geographically closest CBS.
% The CP has access to offline estimates of  content popularity in each association area. 
 Based on offline estimates of  content popularity in each association area, the CP can calculate the potential savings  per CBS when caching each content file. %for each content file   if this file is cached. 
 The optimal strategy for the CP is to  cache exactly those files for which the savings exceed the corresponding cache leasing costs. 
 In this case the MNO associates users independently of their requests and CP decisions. Note that, for zero costs and unit file size, it is optimal to place the locally Most Popular Content in each CBS.
 
\emph{\Opt:}  A different MNO  association policy  allows for more user requests to be served by the caches. If users can be associated to any single wireless node among those close enough (in contrast to a predetermined association, e.g.\ \Closest) each user has potential access to the set of files  cached at all covering stations. It is then beneficial  for the CP to cache  different sets of content in neighboring nodes. To maximize those benefits, the MNO should associate users to CBSs in such a way that content requests are matched to the cached content. A special case for such user association is used for FemtoCaching \cite{Caire2013}.

%The aim of this work is to find the optimal cache leasing and content placement strategy for any network topology, geographic content popularity and user association policy which appropriately adapts to the installed   cache sizes and pricing schemes.

The contributions of this work are summarized as follows:
\begin{compactitem}
\item We introduce a  model whose elements include cache leasing at CBSs, content placement decisions and  user association.
\item We formulate a mixed integer Network Utility Maximization (NUM) problem that aims to maximize the CP's benefits  given the user association policy. The leasing  and caching decisions are discrete, taken by the CP. The MNO association decision variables are fractional. 
\item As solution technique, we introduce Benders decomposition of the NUM which converges to the global optimum. One of its main advantages, aside optimality, is that it allows  the separation of the user association problem from the cache leasing and  content placement problems in the solution process. This separation of the solution process into two subproblems solved iteratively by the two actors is reminiscent of the work by Kelly~\cite{Kelly1997}. Our solution  is completely original for solving NUM edge-caching problems with non-linear utilities.
\item We provide extensive evaluation of the optimal leasing, placement and pricing strategies for linear and concave objective functions. We further show the benefits in hit probability, when the 
MNO makes optimal association decisions taking content placement into account, compared to content agnostic strategies, like the closest node association policy.
\end{compactitem}

The remainder of this paper is organized as follows: In Section~\ref{section:literature}, we survey relevant literature to our problem. The system model is developed in Section~\ref{section:model} where we also state the general mixed integer NUM problem. The general solution based on Benders decomposition is provided in Section~\ref{section:solution}. Special formulations of the problem with linear or separable concave objective functions as well as different association policies (\Closest, \Opt) are presented in Section~\ref{section:cases}. The Section also includes a discussion on how to include wireless resource sharing in the 
problem formulation. Extensive numerical evaluation of these problems and analysis of the findings is given in Section~\ref{section:simulations}. Finally,  Section~\ref{section:conclusion} concludes our work.

%A common way to measure the reduction in data center traffic is the hit-ratio the CP can achieve through %The CP needs to weigh the monetary savings obtained by traffic reduction at the data center against the investment costs incurring through cache leasing. %is valued with a certain amount of money. In other words, the savings function is a linear function on the hit ratio of a content placement decision.
% In this work, we find the optimal strategy for the trade-off between these two aspects.

%The cache leasing and content placement decisions are renewed periodically, for example, several times per day.  The decisions are based on statistical knowledge about the popularity of the content, which varies in time as well as geographically. For example, the popularity of sports related videos around sport stadiums might be predictably high immediately before a sport event while certain music content might be more requested on and near university campuses -- after lecture hours. 
%The cache leasing as well as the content placement decisions are then considered static during one time period, making data center as well as backhaul capacity available for cache-unrelated traffic. This can be particularly beneficial when the data transfer for a new cache placement is performed in off-peak hours.

\section{Related Literature}
\label{section:literature}

There is extensive literature on the advantages of caching in wireless networks. However,  just few works  treat the joint  
leasing, placement and routing problem,  none of which finds 
the exact optimal solution  until now. %Additionally, no other
%work explicitly differentiates between MNO and CP actions within the solution process.
 %Thus, this model does not actually balance the user load.
 
Shanmugam et al.~\cite{Caire2013} place content into caches of small cells to minimize the network delay. Decisions are taken based on popularity data. The users are  associated  to any covering station without taking resource constraints of the wireless nodes into consideration.
 In \cite{Blaszczyszyn2014}, B{\l}aszczyszyn and Giovanidis develop a   content placement policy which maximizes the hit ratio.
 This general  probabilistic solution is not tailored to specific network topologies. 
According to Ba\c{s}tu{\u{g}}, Bennis and Debbah~\cite{Bastug2015}, the globally most popular files are stored in wireless caches. Users are associated to the closest base station, not knowing if the requested content is stored in the cache or not. % The number of users associated to different base stations is not taken into account.
Poularakis, Iosifidis, and  Tassiulas maximize in \cite{Poularakis2014b}  the hit ratio by means of integer optimization. They introduce a bandwidth constraint limiting   the number of users that can be connected to each cellular station.  
A polynomial-time approximation scheme is also provided.
%In their work, an approximation scheme for the maximization of the hit ratio is provided. % The proposed approximation algorithm needs to run in a centralized manner.
%\cite{Poularakis2016} proposes a business model in which network operators pay residential internet users in order to use their WiFi bandwidth and memory at the access points  as caching infrastructure. The authors find an optimal pricing and content placement policy with which mobile data traffic can be offloaded to the local caches.
Deghan et al.~\cite{Dehghan2015}  develop an approximation algorithm for the joint content placement and user association problem minimizing network delay. Their model controls if users are routed through a cached or an uncached path. The users on a cached path are always associated to the closest cache storing the content.
Naveen et al.~\cite{Naveen2015} provide an optimal  placement and user association scheme with fractional content placement. In \cite{Krolikowski2017}, the authors solve  a NUM user association problem assuming that content is already placed in caches.

Considering cache leasing, Poularakis et al.~\cite{Poularakis2016} propose a business model where residential internet users lease part of their wireless bandwidth and storage capacity to the MNO in exchange for financial reimbursement. A joint optimization of incentive, content placement and routing policies provides  offloading of backhaul traffic to local caches. However, their Lagrangian based solution does not converge to the global optimum due to weak duality (see p.~143 in \cite{Bertsimas2005}) and other solution techniques are necessary to solve the problem optimally. 
The idea of sharing backbone cache memory among different CPs is introduced by Araldo, Dan and Rossi~\cite{Araldo2016}. The partitioning of the caches remains under the control of the internet service provider, while the CPs are allowed to establish secure connections between caches and users. Another work on the topic by Douros et al.\ is \cite{Douros2017}. Paschos et al.~\cite{Paschos2016} discuss the roles of MNOs, CPs and users as well as practical aspects of wireless caching such as the compilation of popularity data and the limitations arising from transmission encryption. 

Related to the solution approach, Bekta\c{s} et al.~\cite{Bektacs2008} use Benders decomposition for joint placement and routing problems that have linear utilities and binary variables  in the context of Content Delivery Networks. Their work uses a similar technique as in our work. The main difference is that we treat a wireless network and
we use Generalised Benders’ decomposition for non-linear NUM problems. Our problem here is mixed integer with continuous association variables. 
The decomposition  in our paper has a natural business interpretation.  
%In our decentralized approach, we assume the content placement fixed for a longer period of time. This  requires a  load balancing scheme in its own right that is not covered by this work. %Furthermore, our work introduces novel solution methods necessary for the analysis.

\section{Problem Statement and System Model}

\label{section:model}

\subsection{Problem statement}
\label{section:problem}

The objective of the CP is to lease cache memory at the CBSs and place content into it such that the relation of its expected savings to the leasing cost is optimal. 
The savings are given by the function $\savings(\cdot)$ that takes as input the user  association vector $\yvec^{\uapol}(\xvec)$ where $\xvec$ is the content placement action and $\uapol$ is the MNO's association policy.
The leasing costs at each CBS $\bs$ are the  product of leased units $\z_{\bs}$ times the price per unit $\price_{\bs}$
that is set by the MNO. An additional  fee for the appropriate user association and content delivery can be included. %The expected savings depend the MNO's known association policy $\uapol$.
Formally, the CP seeks a feasible tuple of vectors $(\xvec,\zvec)\in\XZ$ that maximizes the objective function $\savings(\yvec^{\uapol}(\xvec)) - \sum_{\bs\in\BS} \price_{\bs}\z_{\bs}$.
 %Given association policy $\uapol$, the CP has the following general cache leasing and content placement (genCLCP-$\uapol$) problem to solve.
%\begin{align*}
%(\text{genCLCP-}\uapol) \qquad \max_{(\xvec,\zvec)\in\XZ} \quad  \savings(\yvec^{\uapol}(\xvec)) - \sum_{\bs\in\BS} \price_{\bs}\z_{\bs}
%\end{align*}
%In general, this problem is hard since it is a non-linear integer program. 

%In the previous section we have shown that for both association policies  \Closest  and \Opt that we examine in this work, the MNO's association $\yvec^{\uapol}(\xvec)$ is an optimal solution of the problem UA-$\uapol$.
The CP's Cache Leasing and Content Placement problem  (CLCP) can be formulated as the Non-Linear  Mixed-Integer Problem (NLMIP)
\begin{align*}
(\text{CLCP}) & \max_{\substack{(\xvec,\zvec)\in\XZ\\ \yvec\in\Y^{\uapol}}} && \savings(\yvec) - \sum_{\bs\in\BS} \price_{\bs}\z_{\bs}&\\
& \qquad\text{s.\ t.} && 
  y_{\bs,\region,f}  \leq \numUT_{\region, f} \x_{\bs,f}, \fa \bs, \region, f,
\end{align*}
where $\bs$ is a CBS, $\region$ is a planar network region and $f$ is a data file.
All components of this problem will be formally presented and explained in this section.

%The problem class NLMIP consists of hard problems. %Thus, problem-specific solution techniques are required to make CLCP tractable.

%\begin{align*}
%(\text{CLCP}) & \max_{\substack{\zvec\in\Z_{\geq 0}^{\lcard \BS \rcard}\\ \xvec\in\lbrace 0,1\rbrace^{\lcard \BS \rcard\lcard \Files \rcard}\\ \yvec\in\R^{\sum_{\bs\in\BS}\bsRegions(\bs)\lvert \Files \rvert}}} && \savings(\yvec) - \sum_{\bs\in\BS} \price_{\bs}\z_{\bs}&\\
%& \qquad\text{s.\ t.} && \z_{\bs} \leq \mem_{\bs} &\fa \bs\in\BS,\\ 
%&&& \sum_{f\in\Files} \fsize_{f} \x_{\bs, f} \leq \z_{\bs} &\fa \bs\in\BS, \\
%&&& y_{\bs,\region,f}  \leq \numUT_{\region, f} \x_{\bs,f}, &\fa \bs\in\BS, \region\in\bsRegions(\bs), f\in\Files,\\
%&&& \sum_{\bs\in \coverBS(\region)} \sum_{f\in \Files} y_{\bs,\region,f}  \leq \numUT_{\region, f}, &\fa \region\in\Regions, f\in\Files.
%\end{align*}

\subsection{Cache Leasing and Content Placement}
We consider  a cellular communications network with a finite set $\BS$ of CBSs.  % Additionally, there is a Master Base Station (MBS) $\bs_{0}$ which provides coverage to  the entire area of the network. %The set of CBSs and the MBS together are called the base stations (BSs). %The set of all BSs is denoted by $\BS_{+0} = \BS \cup \lbrace \bs_{0} \rbrace$. 
Each CBS $\bs$ is equipped with $\mem_{\bs}$ memory units of size $\fsize_{\text{MU}}$ (in MBytes, e.g.\  1000) which the CP can lease. Denoting the decision variable of how many cache units to lease (CP-1) at $\bs$ by $\z_{\bs}\in\Z_{\geq 0}$, the bounded availability of memory gives the constraint set
\begin{align}
\label{const1}
\z_{\bs} \leq \mem_{\bs} \fa \bs\in\BS.
\end{align} 
The vector of the cache leasing variables is $\zvec = (\z_{\bs})_{\bs\in\BS}$. 

Having leased cache space at the CBSs, the CP places content from a finite object catalog $\Files$ into the caches (CP-2). The decision to store content  $f$  in the cache of $\bs$ will set the variable $\x_{\bs,f}$ to $1$, otherwise  $\x_{\bs,f} = 0$. The vector of content placement variables is $\xvec = (x_{\bs,f})_{\bs\in\BS,f\in\Files}$. Each file $f\in\Files$ has a given file size  $\fsize_{f}$ (in MBytes), and all file-sizes are known. The limited capacity of the leased cache space gives the second constraint set
\begin{align}
\label{const2}
\sum_{f\in\Files} \fsize_{f} \x_{\bs, f} \leq \fsize_{\text{MU}} \z_{\bs} \fa \bs\in\BS.
\end{align}
For convenience, we define the set of feasible tuples of leasing and placement vectors as
\begin{align*}
\XZ \coloneqq \big\{ (\xvec, \zvec) \in \lbrace 0,1\rbrace^{\lcard \BS \rcard\lcard \Files \rcard} \times \Z_{\geq 0}^{\lcard \BS \rcard}\bigm\mid \eqref{const1}, \eqref{const2}\big\}.
\end{align*}

\subsection{Wireless Environment and User Association}
\label{sec:userassoc}

\emph{Coverage Cells}:
Our communications model is the following: 
Each CBS has a planar 2D coverage cell. Users  covered by  a CBS receive a radio signal strong enough to be potentially associated to it. Coverage cells may overlap, thus offering the users multiple options for service from covering CBSs. However, we do not allow simultaneous service by more than one station, i.e.\ cooperative service is not possible.

\emph{Network Regions}:
%The network area is the area of all positions  covered by at least one CBS. 
%The policy $\uapol$  by which users are associated to one of the covering CBSs is decided by the MNO. Depending on $\uapol$,
The network area is partitioned into a set of regions. %The regions represent different  radio conditions in the network as well as the consequences of the MNO's user association policy $\uapol$. 
All positions in each region are assumed to experience the same radio conditions with respect to fading and interference. Furthermore, the MNO has a user association policy $\uapol$ that allows for users in region $\region$ to potentially be associated to any  CBS in $\coverBS(\region)\subseteq\BS$, $\lcard \coverBS(\region)\rcard \geq 1$. (a) With the traditional policy that associates users to the closest station, there can be several covering CBSs for region $\region$, but $\coverBS(\region)$ is the  set consisting of only the closest station. (b) For the \Opt policy, to be defined later, $\coverBS(\region)$ is the set of covering CBSs. %There can, however, be several regions with different radio conditions that associate potentially to the same subset of CBSs.
In general, for policy $\uapol$, the set of regions is denoted by $\Regions^{\uapol}$ (see Figure~\ref{fig:simpleExNetwork} for an example).  %For each $\bs\in\coverBS(\region)$, the association of  a user requesting file $f$ in $\region$ has a  CBS- as well as file-specific weight $\w_{\bs,\region,f}\geq 0$. Possible interpretations of these weights will be given later.

\emph{Content Popularity}:
 For each region $\region\in\Regions^{\uapol}$ and each content $f\in\Files$, the expected number of users in $\region$ requesting $f$ is considered to be known, measured or estimated by some process we do not consider here. It is denoted by $\numUT_{\region,f}$.  The content popularity vector  is  $\mathbf{\numUT} = (\numUT_{\region, f})_{\region\in\Regions^{\uapol},f\in\Files}$.  %This information is available in particular to the CP.
%In the context of this work, we concentrate on  \emph{cache-related} traffic, i.e.\ the users which are associated by the MNO with CBSs which have the requested content cached.

\emph{User Association Variables and Constraints}:
%The CP takes the leasing and placement decisions expecting its users to be associated with the CBSs.
In order to make optimal  decisions, the CP has two types of information at its disposal: the  popularity vector $\mathbf{\numUT}$  %= (\numUT_{\region, f})_{\region\in\Regions,f\in\Files}$ 
and   the MNO's user association policy $\uapol$. %$%\Regions$ is the set of regions which partition the network area.
Knowledge of $\mathbf{\numUT}$ and $\uapol$ allows the CP to take decisions based on the expected association of users with CBSs. In the context of this work, we are only interested in \emph{cache-hit} traffic, i.e.\ the traffic of users who find their request cached at the CBS they are associated to. The  association vector of \emph{cache-hit}  users  to the CBSs is $\yvec = (\y_{\bs,\region,f})_{\bs\in\BS, \region\in\bsRegions^{\uapol}(\bs), f\in\Files}$, where $\y_{\bs,\region,f}$ represents the expected  user traffic from region $\region$ requesting content $f$ and associated with CBS $\bs$. $\bsRegions^{\uapol}(\bs)$ is the subset of regions whose users can potentially be associated to $\bs$ according to $\uapol$. The vector $\yvec$ has fractional non-negative entries.

User assignment is unique in the sense that an association to two or more CBSs is not allowed. The total population $\numUT_{\region, f}$ is  distributed among the  CBSs $\coverBS(\region)$, and  some of it is potentially not associated to any CBS at all. Thus,
 \begin{align}
 \label{const4}
\sum_{\bs\in \coverBS(\region)}  y_{\bs,\region,f}  \leq \numUT_{\region, f}, \fa \region\in\Regions^{\uapol}, f\in\Files.
\end{align}
This constraint allows for possible splitting of the population $\numUT_{\region, f}$ among the CBSs in $\coverBS(\region)$.
The set of assignment vectors feasible to this constraint set is denoted by
\begin{align*}
\Y^{\uapol} \coloneqq \big\{ \yvec\in\R_{\geq 0}^{\sum_{\bs\in\BS}\lvert \bsRegions^{\uapol}(\bs) \rvert \lvert \Files \rvert} \bigm\mid \eqref{const4}\big\rbrace.
\end{align*}

 %The success of the leasing and placement decisions by the CP depends on the association decisions by the MNO: A user associated to a CBS storing the requested content is a cache hit, all others are cache misses. 
Since we are only interested in cache-hit traffic, $\y_{\bs,\region,f}$ can only be nonzero if $\x_{\bs,f} = 1$, i.e.\ if object $f$ is cached in  station $\bs$. Since no more than the total population requesting content $f$ in $\region$ can be included in $\y_{\bs,\region,f}$,  the following constraint set is valid:
 \begin{align}
 \label{const3}
  y_{\bs,\region,f}  \leq \numUT_{\region, f} \x_{\bs,f}, \fa \bs\in\BS, \region\in\bsRegions^{\uapol}(\bs), f\in\Files.
\end{align}
This constraint set is very important since it couples MNO association variables with CP cache placement decisions.

%\subsection{Savings and User Association Policies}

\subsection{CP savings}
\label{section:savings}

The CP uses the general \emph{savings function} $\savings(\cdot)$ to measure user association $\yvec$. This function represents the savings (in \euro{}) obtained when users are associated with caches that store the requested content, thus avoiding use of its data centers. 

In this paper, we solve  CLCP  for any  monotonously increasing, continuously differentiable and concave savings function.
Two choices for $\savings(\cdot)$ are particularly of interest: 
\begin{compactenum}
\item[i)] In case that the CP is solely interested in maximizing the hit ratio, it can choose $\savings(\cdot)$ as a \emph{linear} function. 
\item[ii)] Choosing $\savings(\cdot)$ as the sum of \emph{strictly concave} functions (one function per CBS), the CP can include  aspects such as soft resource  requirements and load-balancing. Additional communication conditions (e.g.\ fading and interference) can also be included in  $\savings$. 
\end{compactenum}
 The discussion over  particular choices of $\savings(\cdot)$, which result in problems with different objectives, is postponed to Section~\ref{section:cases}.

%Three choices for $\savings(\cdot)$ are particularly interesting:
%\begin{compactitem}
%\item Linear function $\savings(\yvec) = \text{const} \sum_{\bs\in\BS}\sum_{\region\in\Regions}\sum_{f\in\Files} \y_{\bs,\region,f}$ where const  are the savings per user in \euro{}. 
%\end{compactitem}
% 

\subsection{MNO policy}
\label{section:policy}
%The way   the MNO    associates users to  CBSs  determines  the association vector $\yvec$, and  depends on the MNO's user association policy $\uapol$.  Examples for such policies are:
%\begin{compactenum}
%\item[a) $\Opt$:] Association maximizing the CP's savings function.
%\item[b) $\Closest$:] Association to the closest covering CBS.% independent of content placement.
%\end{compactenum}
%Observe that, due to \eqref{const3},  cache-hit user association $\yvec$ always depends on the placement vector $\xvec$ in all cases. The resulting association vector is  denoted by $\yvec^{\uapol}(\xvec)$. 
%In case $\uapol = \Opt$, the MNO fully cooperates with the CP in the sense that it always adapts its association vector $\yvec$ to the  placement $\xvec$ such that the CP's savings function $\savings$ is maximized. This is achieved by splitting traffic among CBSs given multi-coverage.
%In contrast,   if $\uapol = \Closest$,  association actions are independent of placement $\xvec$, but the entries of the association vector $\yvec$ are positive only for  users that find their content cached.

The way   the MNO    associates users to  CBSs  determines  the association vector $\yvec$, and  depends on the MNO's user association policy $\uapol$.  The  policies considered in this work are:
\begin{compactenum}
\item[1) $\Closest$:] Association to the closest covering CBS.
\item[2) $\Opt$:] Association maximizing the CP's savings function.% independent of content placement.
\end{compactenum}
Observe that \eqref{const3} implies that  cache-hits (vector $\yvec$) depend on the placement $\xvec$ in all cases. The resulting association vector is  denoted by $\yvec^{\uapol}(\xvec)$. 
If $\uapol = \Closest$, the association entry $\y_{\bs,\region,f}$ is positive only for  users that find their content cached. However, association actions do not  depend on placement $\xvec$. 

On the other hand, if $\uapol = \Opt$, the MNO fully cooperates with the CP in the sense that it always adapts its association vector $\yvec$ to the  placement $\xvec$ such that the CP's savings function $\savings$ is maximized. This is achieved by splitting traffic among CBSs given multi-coverage.

%Having determined  savings function $\savings$ and given  policy $\Pi$, the CP has the cache leasing and content placement problem (CLCP) to solve:

\begin{figure}
\center
\includegraphics[width=0.7\columnwidth]{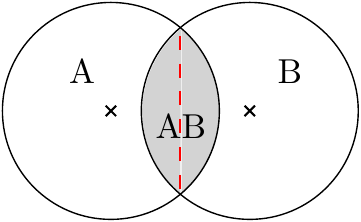}
\caption{In case $\uapol = \Opt$, there are three regions A, B and AB. Users  in region A and B can only be associated to their  uniquely covering CBSs, respecively. Users in region AB can potentially be associated to any of the two CBSs. If $\uapol = \Closest$ (dashed line), there are two regions: A and the left part of AB contain traffic entirely associated to the left CBS, B and the right part of AB contain traffic belonging to the right CBS. }
\label{fig:simpleExNetwork}
\end{figure}

%In the context of this work, we specifically will examine and compare the policies $\uapol = \Opt$ and $\uapol = \Closest$. For the two policies, the set of association regions $\Regions^{\uapol}$ is different as explained in \ref{sec:userassoc} (User Association). We illustrate this difference in the example of Figure~\ref{fig:simpleExNetwork}. There are three regions for the $\Opt$ policy: A, AB, and B, following from the coverage overlaps. For $\uapol = \Closest$, there are only two regions, exactly one per CBS,  defined by the intersection of the coverage cell 
%with its Voronoi cell. In both cases, the regions are disjoint.
For the two policies $\uapol = \Opt$ and $\uapol = \Closest$, the set of association regions $\Regions^{\uapol}$ is different as explained in \ref{sec:userassoc} (User Association). We illustrate this difference in the example of Figure~\ref{fig:simpleExNetwork}. 
For both policies, the association vector is the optimal solution to the User Association  problem 
\begin{align*}
(\text{UA-}\uapol) &&\yvec^{\Pi}(\xvec) =& \argmax_{\yvec \in\Y^{\uapol}} &&\savings(\yvec)\\
 &&& \qquad \text{s.t.}  &&\eqref{const3}.
\end{align*}
This problem is  convex, thus always tractable. 

For \Closest, the  association vector $\yvec^{\Closest}(\xvec)$ follows immediately since $\y_{\bs,\region,f} = \numUT_{\region,f}$ if $\x_{\bs,f} = 1$ and  $\y_{\bs,\region,f} = 0$ otherwise. These values also are the optimum of UA-\Closest, since $\savings$ is strictly increasing.  For $\Opt$, the solution   of $\text{UA-}\Opt$ can be found by convex programming methods.

%Detailed analysis for example savings functions follows in Section~\ref{section:cases}.

%The association policy that guarantees the optimal solution of the UA is called $\textsc{Opt}$. In this paper, we mostly focus on the case in which  the MNO offers the $\textsc{Opt}$ policy.

%We assume in the context of this work that the interests of the CP and the MNO align to the extent that offloading of the CP's user traffic to caches is equally beneficial for both. Given a placement $\xvec$, $\yvec^{\textsc{Opt}}(\xvec)$ is optimal both for the MNO with its metric $\widehat{\savings}(\cdot)$ and the CP with its metric $\savings(\cdot)$ which translates the UA vector $\yvec$ into the savings the CP has from easing traffic on its backhaul infrastructure and the improvement of the service to its users (in \euro{}). The CP, however, has to take the costs for the leasing of cache memory into account: For the use of every chunk of memory at CBS $\bs$, the CP has to pay $\price_{\bs}$ (in \euro{}). 

\section{Solution}
\label{section:solution}

The high complexity of  CLCP  arises from the fact that it is a mixed-integer problem with non-linear objective. The solution technique that resolves this problem is Generalized Benders decomposition by Schrijver~\cite{Schrijver1986} and Geoffrion~\cite{Geoffrion72} which converges to the global optimum.  As in the work of Elwalid, Mitra and Wang \cite{ElwalidMW06} who also use this technique in a different setting, we  solve the non-linear part and the integer part  separately.  CLCP is a generalization of the Helper Decision Problem proved to be NP-complete in \cite{Caire2013} even in the case of a linear savings function. Thus, our solution algorithm cannot be polynomial unless $\text{P} = \text{NP}$.  However,  a state-of-the-art MIP solver can be used for the solution.  Its  performance  is shown  in Section~\ref{section:simulations}.
%  we decompose  a problem involving both integer  and fractional variables with a nonlinear objective function and solve it iteratively.
%Additionally, we have to deal with  constraints \eqref{const4} on the continuous subproblem that are implied in the domain $\mathbb{Y}$. 
In what follows, we  give an overview over Benders decomposition applied to CLCP.

%Consider the  optimization problem
%\begin{align*}
% \text{(P)} && \max_{\substack{(\xvec,\zvec)\in\XZ\\ \yvec\in\Y}} & \savings(\yvec) - \sum_{\bs\in\BS} \price_{\bs}\z_{\bs}&\nonumber\\
%&& & y_{\bs,\region,f}  \leq \numUT_{\region, f} \x_{\bs,f}, \fa \bs\in\BS, \region\in\bsRegions(\bs), f\in\Files, 
%\end{align*}
%where $\mathbf{X}$ and $\mathbf{Z}$ are discrete and finite, $\Y$ is a convex set and $\savings$ is monotonously increasing, continuously differentiable and concave. For convenience, the vectors $\xvec$, $\zvec$ and $\yvec$ are defined as in Section~\ref{section:model}. 

CLCP can be decomposed  into two problems called  Master  and  Slave. Master decides about cache leasing and content placement. Slave  computes the optimal user association for a fixed content placement. We obtain
\begin{align*}
(\text{Master}) &&\max_{(\xvec,\zvec)\in\XZ} &\qquad\savings(\yvec(\xvec)) - \sum_{\bs\in\BS} \price_{\bs}\z_{\bs}, 
\end{align*}
where $\savings(\yvec(\xvec))$ is the objective value of
\begin{align*}
(\text{Slave}) &&\yvec(\xvec) = \argmax_{\yvec\in\Y^{\uapol}} &\qquad\savings(\yvec)  \\
&& \text{s.\ t.} &\qquad\eqref{const3}.
\end{align*}
Note that the Master problem can be treated by the CP,  the Slave by the MNO.
$\XZ$ is discrete and finite and $\Y^{\uapol}$ is compact and convex. %\footnote{Since the solution process is the same both for $\uapol = \Closest$ and $\uapol = \Opt$, we omit the index $\uapol$ here and in the following and write $\Y$ for $\Y^{\uapol}$ as well as $\Regions$ for $\Regions^{\uapol}$ and UA for UA-$\uapol$.} 
 Slave is   the UA-$\uapol $ problem from Section~\ref{section:policy} which is  generally non-linear. Thus,  Master cannot be solved directly. Generalized Benders decomposition deals with this problem by solving a sequence of Slave problems for different values of $\xvec$ (and $\zvec$). The  solutions to Slave are  used to construct (linear) Benders cuts that define approximations to  Slave, thus linearizing the problem.

Let $\lbrace(\xvec^{t},\zvec^{t}) \in\XZ\mid  t = 1,\ldots,T\rbrace$ be a set of vector tuples feasible to Master for some $T\geq 0$. Let $\yvec^{t}\coloneqq \yvec(\xvec^{t})$ denote a corresponding vector  that optimizes Slave for given $\xvec^{t}$, and $\lagvec^{t}= (\lag^{t}_{\bs,\region,f})_{\bs\in\BS, \region\in\bsRegions^{\uapol}(\bs), f\in\Files}$ be the  vector of Lagrangian multipliers corresponding to the constraints \eqref{const3}.
The existence  of such vector $\yvec^{t}$ follows from the fact that $\numUT_{\region,f}\geq 0$ holds true for all $\region\in\Regions^{\uapol}, f\in\Files$. The computability of $\yvec^{t}$ and $\lagvec^{t}$ are assumed here, and discussed for special cases in Section~\ref{section:cases}. 

Slave is a convex problem. Thus, the duality theorem of convex programming implies
\begin{align*}
\savings(\yvec(\xvec)) \leq  &\qquad \savings(\yvec^{t}) \quad + \\  & \sum_{\bs\in\BS}\sum_{\region\in\bsRegions^{\uapol}(\bs)}\sum_{f\in\Files} \lag^{t}_{\bs,\region,f}(\numUT_{\region, f} \x_{\bs,f} - y^{t}_{\bs,\region,f})
\end{align*}
for all feasible vectors $\xvec$. This upper bound to Slave is called \emph{Benders cut}. Reformulated, we get
\begin{align}
\label{benders-upper-bound}
\savings(\yvec(\xvec)) \leq \Gamma^{t}  + (\bendersvec^{t})^{\prime}\xvec,
\end{align}
where
\begin{align*}
\Gamma^{t} \coloneqq \savings(\yvec^{t}) - \sum_{\bs\in\BS}\sum_{\region\in\bsRegions^{\uapol}(\bs)}\sum_{f\in\Files} \lag^{t}_{\bs,\region,f} y^{t}_{\bs,\region,f}
\end{align*}
and $(\bendersvec^{t})^{\prime}$ is the transpose of $\bendersvec^{t} = (\benders^{t}_{\bs,f})_{\bs\in\BS, f\in\Files}$ with
$
\benders^{t}_{\bs,f} = \sum_{\region\in\bsRegions^{\uapol}(\bs)} \lag^{t}_{\bs,\region,f}\numUT_{\region, f}.
$

For a set of $T$ Benders cuts  (for some $T\geq 1$), we obtain an upper bound to the  original problem CLCP by solving  the surrogate IP
\begin{align*}
(\text{SUR-$T$}) & \max_{\substack{(\xvec,\zvec)\in\XZ\\ \gamma \in\R_{\geq 0}}} && \gamma - \sum_{\bs\in\BS} \price_{\bs}\z_{\bs}\\
&\qquad\text{s.\ t.} &&\gamma \leq \Gamma^{t}  + (\bendersvec^{t})^{\prime}\xvec \qquad t=1, \ldots ,T.
\end{align*}
Note that the auxiliary variable $\gamma$,  together with  the Benders cuts,   approximates Slave linearly. This way, SUR-$T$ avoids the non-linearity which creates the difficulty for solving Master. 
The optimal objective value  to the surrogate problem is denoted by $\sur^{T}$.
 The optimal solution  consists of $\xvec^{T+1}$, $\zvec^{T+1}$  and $\gamma^{T+1}$.

Generalized Benders decomposition is an iterative process. We start in step $0$ with an initial feasible tuple of leasing and placement vectors $(\xvec^{0},\zvec^{0})\in\XZ$ and without any Benders cuts. At the start of step $T\geq 0$, we have the current leasing and placement $(\xvec^{T},\zvec^{T})\in\XZ$ and  $T$ Benders cuts. We solve  Slave with input $\xvec^{T}$ to obtain the optimal association vector $\yvec^{T}$. Clearly, since the triple $\xvec^{T}$, $\zvec^{T}$ and $\yvec^{T}$ are feasible  to the  original problem CLCP, its corresponding objective value provides  a lower bound to the optimal value of CLCP. Additionally, we compute the Lagrangian multipliers $\lagvec^{T}$ and the corresponding $(T+1)$th Benders cut \eqref{benders-upper-bound}. With the Benders cut we obtain the surrogate MIP SUR-($T+1$).  Its optimal solution is the feasible leasing and placement vectors $(\xvec^{T+1},\zvec^{T+1})\in\XZ$. The objective value  $\sur^{T+1}$ is an  upper bound to CLCP. This process iterates. At any step $T$, $\sur^{T}$ is the current upper bound (note that $\sur^{T+1}\leq \sur^{T}$ after every step $T$), while the current lower bound is provided by the best found solution $\max_{t\in\{0,\ldots,T\}} \savings(\yvec^{t}) - \sum_{\bs\in\BS} \price_{\bs}\z^{t}_{\bs}$.
 The  process terminates in the globally optimal cache leasing and content placement vectors $\zvec^{\ast}$ and $\xvec^{\ast}$ when the upper and lower bounds coincide.   
Convergence is guaranteed from the proof of Theorem 2.4 in \cite{Geoffrion72} and the fact that the domain $\XZ$ of the Master  is finite.
In every step $T$, an instance of Master needs to be solved. 

For general parameters $\Gamma^{t}$ and $\bendersvec^{t}$, $T=1,\ldots,T$, it is  NP-complete. This can be shown with a reduction from the classic \textsc{Set Cover} problem. State of the art MIP solvers such as CPLEX are, however, capable of solving SUR-$T$ in reasonable runtime. We state without proof (due to space limitations) that the SUR-$T$ problem has an infinite integrality gap (gap between linear relaxation and optimal discrete solution). 
Approximation algorithms for SUR-$T$ are a topic for future research.

\section{Special Cases}
\label{section:cases}

The savings function $\savings$ introduced in Section~\ref{section:savings} maps the user association vector $\yvec$ onto the savings $\savings(\yvec)$ (in \euro{}). %The savings function represents the common interest of the MNO and the CP: They both value a user association which 1) increases the cache-related traffic and b) balances the load among the CBSs.\footnote{In future work, different interests of the MNO and the CP regarding user association will be explored.} 
%The savings function is  monotonously increasing, concave and continuously differentiable. 
In the following, we give examples of specific expressions for $\savings(\cdot)$ and explain what each example implies for the solution. %We also remind the reader that in every iteration of the Benders decomposition, one instance of the slave problem (S) needs to be solved that determines user association maximizing  $\savings(\cdot)$.
%In our examples, the user traffic routed through each CBS is particularly important. 
We denote the cache-hit traffic volume associated to CBS $\bs$ through vector $\yvec$ by
\begin{align*}
\vol_{\bs}(\yvec) = \sum_{\region\in\bsRegions^{\uapol}(\bs)}\sum_{f\in\Files} \y_{\bs,\region,f}.
\end{align*}

\subsection{Linear Savings (case i)}% without Interference, Fading or Resource Constraints}
\label{section:simple-hit-ratio}

As a  first case,  the  cache-hit user traffic  of CBSs is linearly mapped onto monetary benefits for the CP, i.e.\
\begin{align}
\label{savings-hr}
\savings^{\text{L}}(\yvec) = c \sum_{\bs\in\BS} \vol_{\bs}(\yvec),
\end{align} 
where $c$ is the savings (in \euro{}) per cache hit. Since the popularity vector is constant, the savings value is proportional to the hit ratio. The latter is simply calculated by $\savings^{\text{L}}(\yvec) / (c \sum_{\region\in\Regions^{\uapol}}\sum_{f\in\Files} \numUT_{\region,f})$.

With a linear savings function, the slave problem becomes easily tractable for any association policy $\uapol$, including \Closest and \Opt:
%\subsubsection{Solving (S) with savings function \eqref{savings-hr}}
Given a CP vector $\xvec$, the MNO can freely distribute users among the covering CBSs $\coverBS(\region)$ which have $f$ cached. The association to any CBS contributes equally to the savings.  If no  $\bs\in\coverBS(\region)$ caches $f$, then $\y_{\bs,\region,f} = 0$, hence no cache hit  from region $\region$ for file $f$. 

\subsection{Separable Concave Savings (case ii)}
\label{section:constraints}

As a second case, we introduce the sum of strictly concave functions, one per CBS, taking as argument the associated traffic volume. 
The strict concavity of the utility functions implies diminishing returns for user traffic in every CBS. This way, the MNO has the incentive to associate users with underused CBSs while the overuse of CBSs is disincentivized. 
%This type of expression increases fast when the associated traffic is low, thus pushing the optimization algorithm to associate traffic to underused CBSs.
%However, the function has diminishing returns per station, when the volume of users associated further increases, thus making additional association less and less beneficial. 
This choice for $\savings$ can model physical resource limitations on each CBS that prohibit the good service of users when their volume becomes 
very high. As a result,  a type of load balancing among CBSs is enforced.
Observe that in this way, 
 we do not impose a hard constraint on the user traffic but introduce a soft constraint in the objective function instead. 
 
Formally, we define utility functions $\util_{\bs}(\cdot)$ for every $\bs\in\BS$ which are monotonously increasing, strictly concave and continuously increasing. The input of the utility functions is the cache-hit traffic volume at the CBS. The savings function is the sum of all utility functions. We obtain
\begin{align}
\label{utility-savings}
\savings^{\text{C}}(\yvec) = \sum_{\bs\in\BS} \util_{\bs}(\vol_{\bs}(\yvec)).
\end{align}
%In practical terms, the utility functions can be chosen such that, for a given CP, the optial offloading from over- to underloaded CBSs can be achieved. 

The  load  can be balanced among the CBSs by   guaranteeing fairness  with regards to associated volume. %In case that $\util_{\bs}(\cdot) = \log(\cdot)$
Some notions of fairness are max-min, $\alpha-$ and proportional fairness. Each of them is achieved by appropriate choice of the utility functions (see~\cite{Kelly1997, Mo2000}). E.g.\ for proportional fairness, utilities are chosen as (weighted) logarithms. %In our case, max-min fairness and proportional fairness are equivalent.
%By enforcing fairness onto UA, we achieve that there are no underused CBSs. Simultaneously, an overload of CBSs is avoided.

For any association policy $\uapol$, the slave problem with savings function as in \eqref{utility-savings} is a convex problem. Particulary the case $\uapol = \Opt$ was  solved in \cite{Krolikowski2017} using Lagrangian duality. The same method can be applied for $\Closest$ as well. 

\subsection{Weighted Savings}
\label{section:interference}

In the previous section, there was no differentiation between wireless users in the objective function. Here, we introduce weights $\w_{\bs,\region,f}\geq 0$ that are specific to users from region $\region$ requesting content $f$ associated to CBS $\bs$. This generalization allows to include costs and benefits from associating certain user groups to particular stations. The weighted traffic volume at CBS $\bs$ is
\begin{align*}
\vol^{\mathbf{\w}}_{\bs}(\yvec) = \sum_{\region\in\bsRegions^{\uapol}(\bs)}\sum_{f\in\Files} \w_{\bs,\region,f} \y_{\bs,\region,f},
\end{align*}
where $\mathbf{\w}$ is the vector of weights. This volume can be used as argument of the weighted savings function (linear or concave)
\begin{align}
\label{weighted-utility-savings}
\savings^{\text{C},\mathbf{\w}}(\yvec) = \sum_{\bs\in\BS} \util_{\bs}(\vol^{\mathbf{\w}}_{\bs}(\yvec))
\end{align}
where $c$ is the savings (in \euro{}) per weighted cache hit.

\emph{i) Prioritized Caching}:
If the weights $\w_{\bs,\region,f}$ are proportional to the file sizes $\fsize_{f}$, files of larger size that create more burden to the backbone are favorized to be cached. Such cases are also treated in \cite{Neglia2017}. %Another option is to select $\mathbf{\w}$ such that users from certain areas are more likely to find their content cached than users from other regions.

\emph{ii) Network Throughput}:
The weights can represent the 
 downlink throughput  between a user and a CBS. For such a weights choice, the channel quality between $\bs$ and $\region$ is a constant value $\chan_{\bs,\region}$ that depends on a reference distance and the path loss exponent. The emitted power level  of $\bs$ is denoted by $\pow_{\bs}$ and the  noise level  by $\noise$. Then, the signal-to-interference-plus-noise ratio (SINR)  of users in region $\region$ when associated to covering CBS $\bs$ is 
\begin{align}
\label{eq:sinr}
\sinr_{\region}(\bs) = \pow_{\bs}\chan_{\bs,\region} \bigg(\sum_{\substack{\tilde{\bs} \text{ covers } \region\\ \tilde{\bs}\neq\bs}} \pow_{\tilde{\bs}}\chan_{\tilde{\bs},\region} + \noise\bigg)^{-1},
\end{align}
where we assume that the  interference from CBSs not covering $\region$ is negligible. For the downlink transmission from CBS $\bs$ to region $\region$, the throughput is equal to 
$
\w_{\bs,\region,f} =B\log_{2}(1+\sinr_{\region}(\bs)) \quad \left[\text{in bits/sec}\right]
$
where $B$ [Hz] is a chunk of bandwidth allocated to each served user. The total service bandwidth per CBS is equal to the product of $B$ times the users routed to the CBS.
 Using these weights, CLCP takes into account that it is favorable for a CBS to serve users with good radio conditions in order to use its resources effectively. Such cases are also treated in \cite{Liu2017b}. % The utility functions can be chosen as concave to balance achieved throughput among the CBSs. 
%the resources consumed for the service of each user associated with a CBS. More precisely, when considering downlink communication, they can be the bandwidth necessary to completely transmit file $f$ from $\bs$ to a user in region $\region$  within a given time limit $\tau$.
 %In this way, we can include the channel quality between $\bs$ and $\region$ as a constant value $\chan_{\bs,\region}$ that depends on a reference distance. The emitted power level  of $\bs$ is denoted by $\pow_{\bs}$ and the background noise level  by $\noise$. Then, the signal-to-interference-plus-noise ratio (SINR)  of users in region $\region$ when associated to covering CBS $\bs$ is 
%\begin{align*}
%\sinr_{\region}(\bs) = \pow_{\bs}\chan_{\bs,\region} \bigg(\sum_{\substack{\tilde{\bs} \text{ covers } \region\\ \tilde{\bs}\neq\bs}} \pow_{\tilde{\bs}}\chan_{\tilde{\bs},\region} + \noise\bigg)^{-1},
%\end{align*}
%where we assume that the  interference from CBSs not covering $\region$ is negligible. For the downlink transmission of a file of size $\fsize_{f}$ from CBS $\bs$ to region $\region$ in time~$\tau$, the required bandwidth is equal to $\w_{\bs,\region,f} =\fsize_{f}(\tau\log_{2}(1+\sinr_{\region}(\bs)))^{-1}$. Using this value as weight in the concave utility \eqref{weighted-utility-savings}, the load balancing can be done over the resource 
%consumption, rather than on the count of users. This way, physical communications aspects are taken into account.

We would like to emphasize that the arbitrarily many  levels of channel quality on the coverage area of each station can be introduced by appropriately redefining  the region set $\Regions$   that partitions the network area. The modelling tradeoff is between the precision of communications aspects and the runtime of the optimization process.

%
%
%
%\begin{figure}
%	\includegraphics[width=0.8\linewidth]{regions}
%	\caption{A network   consisting of two CBSs B1 and B2.  The  coverage areas are depicted as the  solid circular lines. The zones of equal channel quality are separated by the dashed circular lines. The resulting regions are denoted by $s_1$ to $s_7$. All users in each region  $s_i$ requesting file $f$ that are associated to CBS $\bs\in\lbrace\text{B1},\text{B2}\rbrace$ cause the same communication costs $\w_{\bs, s_{i}, f}$.}
%	\label{fig:regions}
%\caption{Toy network and communication costs}
%\label{fig:network}
%\end{figure}

\section{Experiments and Numerical Evaluation}
\label{section:simulations}
\subsection{Environment}

We simulate cellular networks  in an urban environment and calculate the optimal cache leasing and content placement for 4 cases: The savings function $\savings$ is i) linear as in \eqref{savings-hr} or ii) the sum of utility functions as in \eqref{utility-savings}. In the latter case all utility functions $\util_{\bs}$ are chosen as the natural logarithm to achieve proportional fairness for the user traffic at the CBSs. For each of the two cases, the MNO's user association policy is a) the MNO-CP cooperative policy \Opt or b) the conventional policy \Closest.  %All files have unit size $\fsize_{f} = 1$. %This implies that the communication cost from every region to each covering CBS are equal and can be normalized as $\w_{\bs,\region,f} = 1$ for all $\bs\in\BS, \region\in\Regions, f\in\Files$. The utility functions $\util_{\bs}$ are chosen as identity functions for all $\bs\in\BS$, which translates to the utility being proportional to the hit ratio as described in Section~\ref{section:simple-hit-ratio}.
In each case, we simulated 100 random sets of CBS positions as a Poisson Point Process (PPP). This means 
that their total number in each run is a random Poisson realization, and their positions are uniformly distributed in the simulation window. The  density of the PPP is  $80 \frac{\text{CBS}}{\text{km}^2}$ for the cases i.a) and i.b)  (linear savings) and $60 \frac{\text{CBS}}{\text{km}^2}$ for the cases ii.a) and ii.b) (log-savings). This implies an average  minimal distance of 56m and 65m between the CBS positions, respectively. For the cases i.a) and i.b), the evaluation window has size  500$\times$500$\text{m}^{2}$, while the cases ii.a) and ii.b) were evaluated in a 300$\times$300$\text{m}^{2}$ window. The expected number of CBSs in the evaluated windows is 20 for  case i) and 5.4 for  case ii). In both cases, a larger area was simulated to avoid edge effects.
The CBSs's coverage radius varies from  40m to 120m. The MNO price per unit size cache memory at all CBSs varies  (0.01\euro{}-2.00\euro{}). The user population is distributed uniformly over the network with a density of 30 users per $\text{km}^{2}$. The total simulated file catalog contains 100 objects.  The content popularity follows the Zipf distribution with parameter 0.6 unless explicitly stated otherwise.  The available cache size from the MNO is  set to the catalog size so that only the pricing influences  cache leasing decisions.

\subsection{Implementation}

All simulations have been performed using a native JAVA simulation environment. User association corresponding to the solution of the slave problem in Section~\ref{section:solution} is entirely done by optimization algorithms that we developped in the lab, outlined in Section~\ref{section:cases}. The surrogate problem SUR-$T$ in Section~\ref{section:solution} is solved using the state of the art mixed-integer problem solver IBM CPLEX 12.7.0 in combination with IBM ILOG CPLEX Optimization Studio. The experiments have been performed on  a machine with a 2.40 GHz 16-core processor and 48 GB RAM.

% In cases in which \eqref{utility-savings} with $\log$ is chosen as savings function, the runtime of the user association algorithm is longer than in the linear case. 

%Multiple calculations of optimal user assignment $\yvec$ for the same content placement $\xvec$ are avoided by retaining the Benders cuts  for the same network (with constant CBS positions, coverage radius and Zipf parameter) for varying prices. %Furthermore, all calculated Benders cuts can be introduced to the surrogate IPs SUR-$T$ to which operate on the same CBS positions, coverage radius and Zipf parameter. 
%This way, the surrogate problem SUR-$T$ approximates  CLCP-$\textsc{Opt}$ better from the start, and the number of iterations of the Benders decomposition is reduced.

\subsection{Results for Linear Savings Function (case i)}

\begin{figure*}[htp]%
\begin{subfigure}{.66\columnwidth}
\includegraphics[width=\columnwidth]{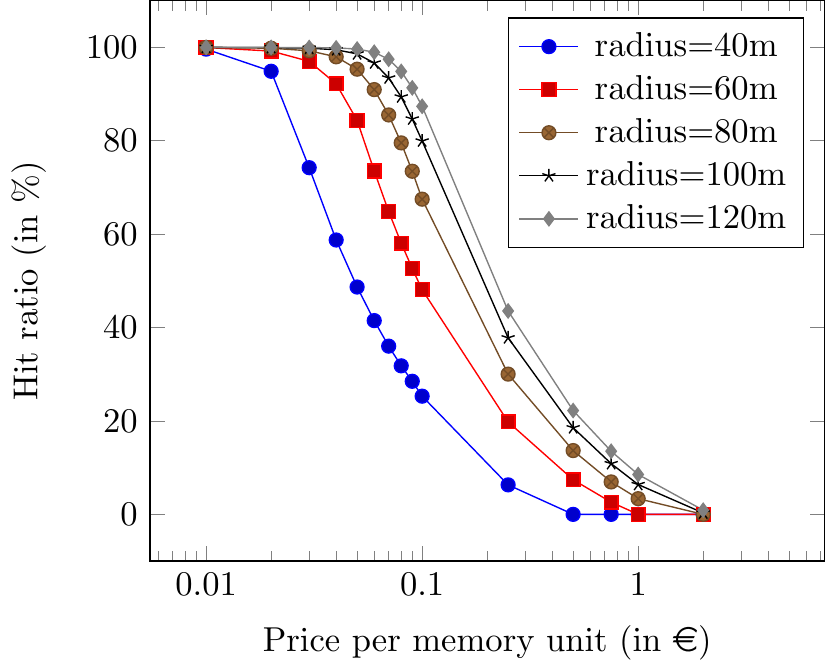}%
\label{fig:lin-distr-hr-price-z06}%
\end{subfigure}\hfill%
\begin{subfigure}{.66\columnwidth}
\includegraphics[width=\columnwidth]{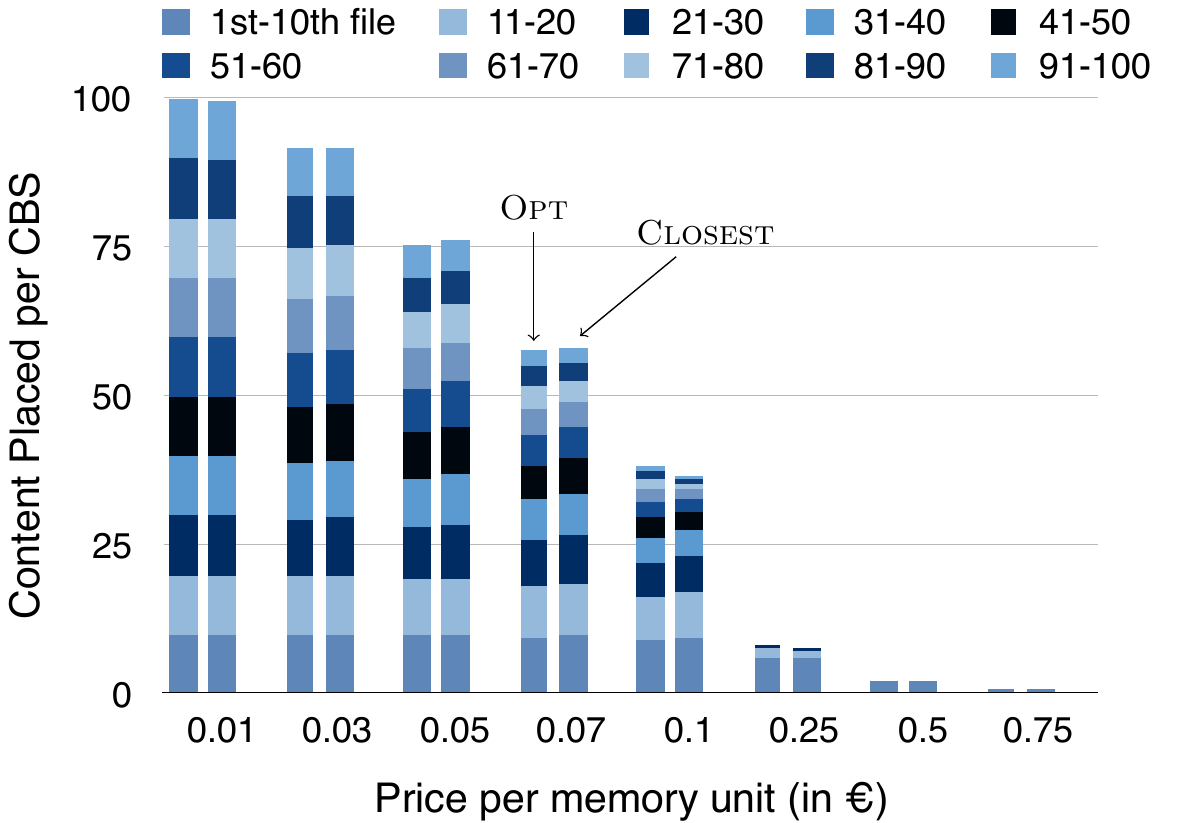}%
\label{fig:lin-both-CP-price}%
\end{subfigure}\hfill%
\begin{subfigure}{.66\columnwidth}
\includegraphics[width=\columnwidth]{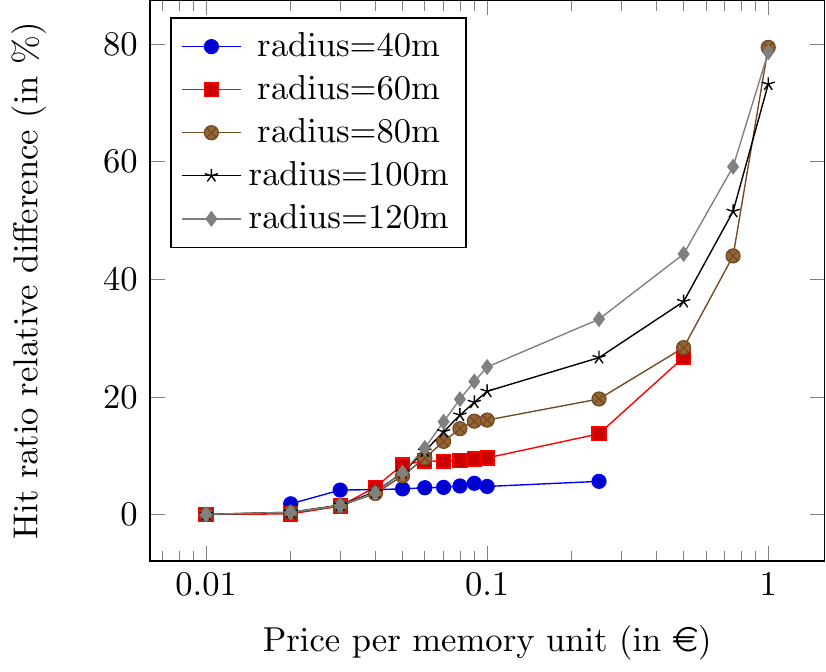}%
\label{fig:lin-diff-hr-price}%
\end{subfigure} 
\caption{Linear savings function. (a): Hit ratio  in relation to price for different coverage radii for case i.a). (b): Cache lease and placement of popular files depending on cache unit price. Each left column represents case i.a), each right column is case i.b). (c): Relative difference between the hit ratio achieved by  $\Opt$ (case i.a) to $\Closest$ (case i.b) over cache unit price for different coverage radii. }
\label{fig:lin1}
\end{figure*}

At first, we present the simulation results for linear savings function. We emphasize  case i.a) which performs \textsc{Opt} user association and compare it with case i.b) \Closest.
Figure~\ref{fig:lin1}(a) %\ref{fig:lin-distr-hr-price-z06}
 illustrates how the hit ratio in case i.a) depends on the price per cache unit for different coverage radii. For all  radii, the hit ratio decreases with increasing prices. With lowest  price (0.01\euro{}/Unit), the CP leases in each CBS the entire memory available, so the hit ratio is $100\%$. As the price increases, the CP leases less units, and the hit ratio is reduced. This happens more quickly in networks with smaller coverage areas because there are less users covered by each CBS  and also  less coverage overlap area. When the price reaches a high level (2\euro{}), the cost from leasing cache memory exceeds the benefit from cache hits and the hit ratio drops to $0\%$ for all  coverage radii. %For prices between $0.03$\euro{} and $0.1$\euro{}, networks with smaller coverage area (e.g.\ 40m and 60m) differ more in hit ratio than networks with larger coverage area (e.g.\ 100m and 120m). 
 The differences between the curves in Figure\ref{fig:lin1}(a) %\ref{fig:lin-distr-hr-price-z06}
 diminish with higher radii where multi-coverage  is already high enough. %Furthermore, with a high coverage radius, increasing the radius  does not further add users to the network since every user in the area is already covered.

The CP's leasing and placement decisions for networks with coverage radius 100m for the  policies \Opt   (i.a)  and \Closest (i.b) are shown in Figure~\ref{fig:lin1}(b). %\ref{fig:lin-both-CP-price}.  
For each price, there are two columns: The lefthand-side column represents the i.a) case,  the righthand-side column  i.b).  The height of each column is the average amount of cache units per CBS which are leased for the respective price. The subdivisions of each column  represent the popularity of the files  stored in the leased memory space: The bottom part are the ten most popular files, the second-to-bottom part  are the files of popularity rank 11 to 20 and so on.
 For the lowest cache price ($0.01$\euro{}), all 100 available units are leased: For both assignment policies, the amount of leased cache memory decreases with increasing price. For the price interval $0.03$-$0.07$\euro{}, \textsc{Opt} uses fewer cache units than $\Closest$. But, as the next Figure~\ref{fig:lin1}(c) %\ref{fig:lin-diff-hr-price} 
 shows, the hit probability by using \textsc{Opt} when the coverage radius is 100m and for the same price interval is $5$-$15\%$ higher than the one achieved by $\Closest$ even with less cached objects. Between $0.1$ and $0.25$\euro{}, \textsc{Opt} uses more cache memory than $\Closest$. However, the corresponding hit ratio is between $20$-$50\%$ higher as well. For all prices, Figure~\ref{fig:lin1}(b) %\ref{fig:lin-both-CP-price} 
  shows that the less popular files are represented more frequently with \textsc{Opt} than with $\Closest$, especially for prices $\geq 0.1$\euro{}. There is more diversity of visible content with the \textsc{Opt} placement.

Figure~\ref{fig:lin1}(c) %\ref{fig:lin-diff-hr-price},
directly compares \textsc{Opt} with \Closest. For all prices the hit ratio achieved by  \textsc{Opt}  (case i.a)  is higher than the one achieved by the $\Closest$ policy (case i.b). In the price interval  $0.06$-$0.50$\euro{}, the relative hit ratio differences are higher when the coverage area of the CBSs is higher. For higher coverage radii (from 80m on) a hit ratio gap of \mbox{$15\%$-$50\%$} can be seen. For higher prices, the $\Closest$ hit ratio is close to 0, therefore the relative differences can become very high.

%The availability of content to a user placed randomly into the covered network area is shown in Figure~\ref{fig:lin-distr-visibile-price}. The 
%
%%f
%\begin{figure}
%\includegraphics{linear-distr-visible-over-price-z06.pdf}
%\caption{Average visible files over cache unit price for different coverage radii for a uniformly randomly selected position in the 500m x 500m network area.}
%\label{fig:lin-distr-visibile-price}
%\end{figure}

\begin{figure*}[htp]%
\begin{subfigure}{.66\columnwidth}
\includegraphics[width=\columnwidth]{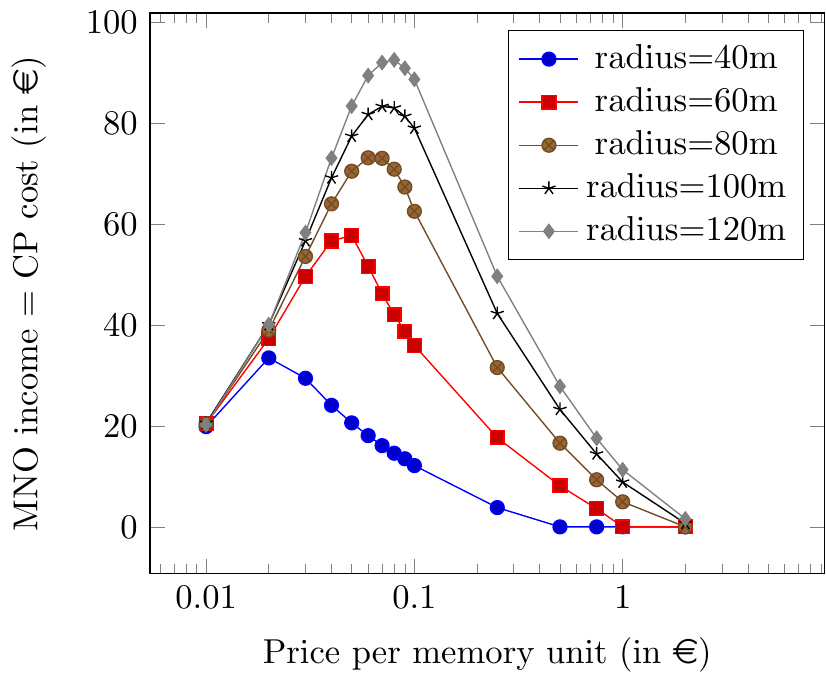}%
\label{fig:lin-closest-income-price}%
\end{subfigure}\hfill%
\begin{subfigure}{.66\columnwidth}
\includegraphics[width=\columnwidth]{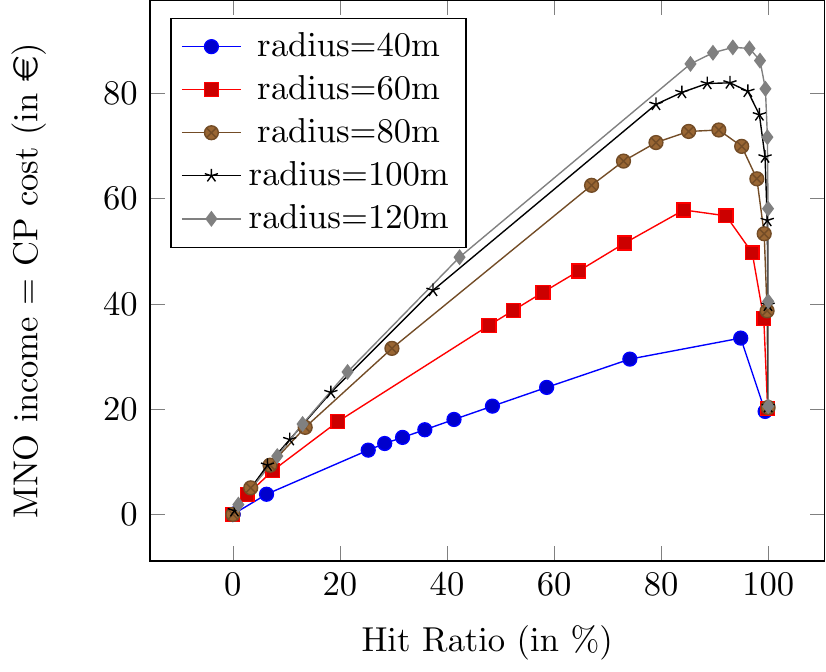}%
\label{fig:lin-distr-hr-income}%
\end{subfigure}\hfill%
\begin{subfigure}{.66\columnwidth}
\includegraphics[width=\columnwidth]{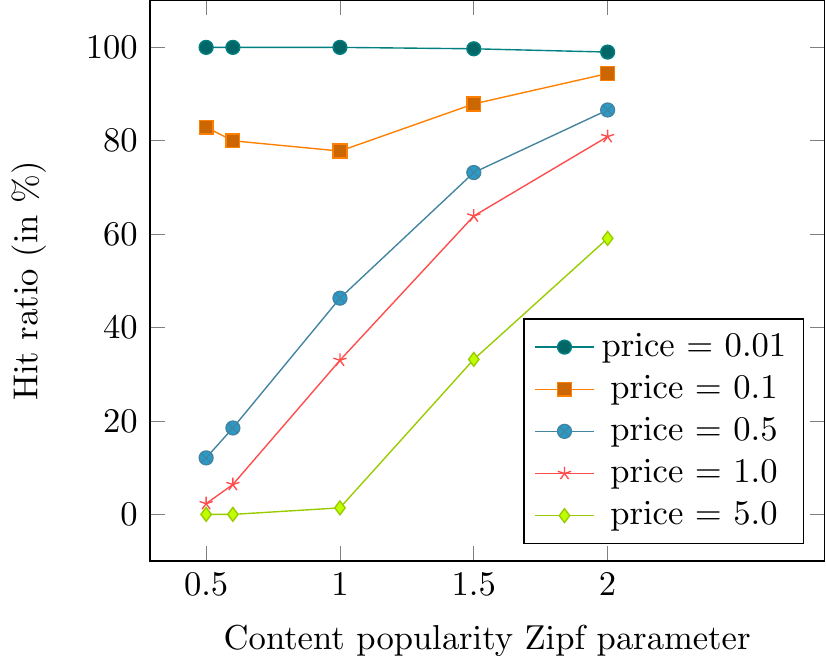}%
\label{fig:lin-distr-hr-zipf}%
\end{subfigure}%
\caption{Linear savings function. (a): Income of MNO in relation to price per cache unit for different coverage radii  in case i.a). (b): MNO income/CP investment over  hit ratio for different coverage radii in case i.a): Achieving the hit ratio on the x-axis results in the MNO income on y-axis. (c): Hit ratio in relation to Zipf parameter for different prices in case i.a).}\label{fig:lin2}
\end{figure*}

While Figure~\ref{fig:lin1} %\ref{fig:lin-distr-hr-price-z06} 
shows the caching benefits for the CP, the costs it has to pay in return to the MNO (in case i.a) are depicted in Figure~\ref{fig:lin2}(a). %\ref{fig:lin-distr-hr-price-z06}. 
The CP costs  equal the MNO income. This amount  can be calculated by multiplying the number of leased cache units with the price per unit. The maximum of the curve  can be clearly identified for each   radius. This is the operational point for the MNO when the latter aims for maximum income. The maxima are higher for larger coverage areas, while the difference in income decreases with increasing radius. Furthermore, the higher the coverage radius, the higher the cache leasing price at which the maximum is achieved. % The operational point of an MNO of a network with 40m coverage radius is at $0.02$\euro{} while for a radius of 120m, it is at $0.08$\euro{}. %For the minimum price, the maximum number of cache units is leased in each case and thus the MNO incomes coincide. The slope towards the operational point is steep  especially for large radius networks since the decrease  in leased cache units decreases slower than the corresponding price increases (see column sizes in Figure~\ref{fig:lin-both-CP-price}). The subsequent decrease comes from the fact that a steep decrease of leased cache units cannot be counterbalanced by incresed prices.

The relation between hit ratio and MNO income can be seen in Figure~\ref{fig:lin2}(b): %\ref{fig:lin-distr-hr-income}: 
The x-axis displays the hit ratio achieved, the y-axis shows the income the MNO  earns. Again, the income is higher in networks with larger coverage area. The maximum income for all simulated networks can be found for an achieved hit ratio between $80$-$90\%$. Conversely, if the CP decides to invest a certain sum, it can maximally achieve the rightmost of the two corresponding hit ratio values under the condition that the MNO chooses the pricing strategy most favorable to the CP.

The experimental results presented until now are based on a Zipf parameter of 0.6. However, a varying Zipf parameter influences the results: Figure~\ref{fig:lin2}(c) %\ref{fig:lin-distr-hr-zipf} 
shows that the higher the price, the lower the hit ratio for any Zipf parameter in case i.a). This is due to the fact that lower price implies more leased units for the CP. For all prices (except the lowest one which achieves a hit ratio of near 100\% throughout), the hit ratio increases with increasing Zipf parameter: With higher Zipf parameter, the population share requesting the most popular files becomes higher, thus  caching  popular files  becomes more profitable and   a higher hit ratio is achieved. Also, for the same price, the leased cache memory is more effective with higher Zipf parameters. %For price 0.01, the curve decreases slightly since the popularity of less popular content becomes so low that caching of it is not even worthwhile for this price.
The  relative difference on hit ratio between cases i.a) (\textsc{Opt}) and i.b) ($\Closest$)  depend on the Zipf parameter as well (not included in the paper). The lower the Zipf parameter, the more pronounced the differences in hit ratio between different cache prices since the benefits from overlapping coverage are bigger when content popularity is more even.

\subsection{Results for Log Savings Function per CBS (case ii)}

\begin{figure*}[t]%
\centering
\begin{subfigure}{.66\columnwidth}
\includegraphics[width=\columnwidth]{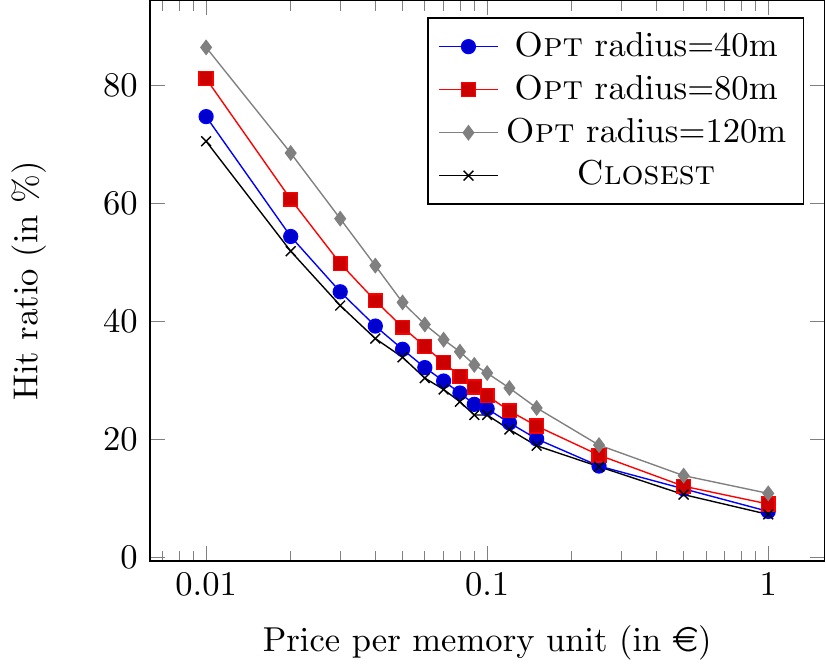}%
\label{fig:log-distr-hr-price-z06}%
\end{subfigure}\hfill%
\begin{subfigure}{.66\columnwidth}
\includegraphics[width=\columnwidth]{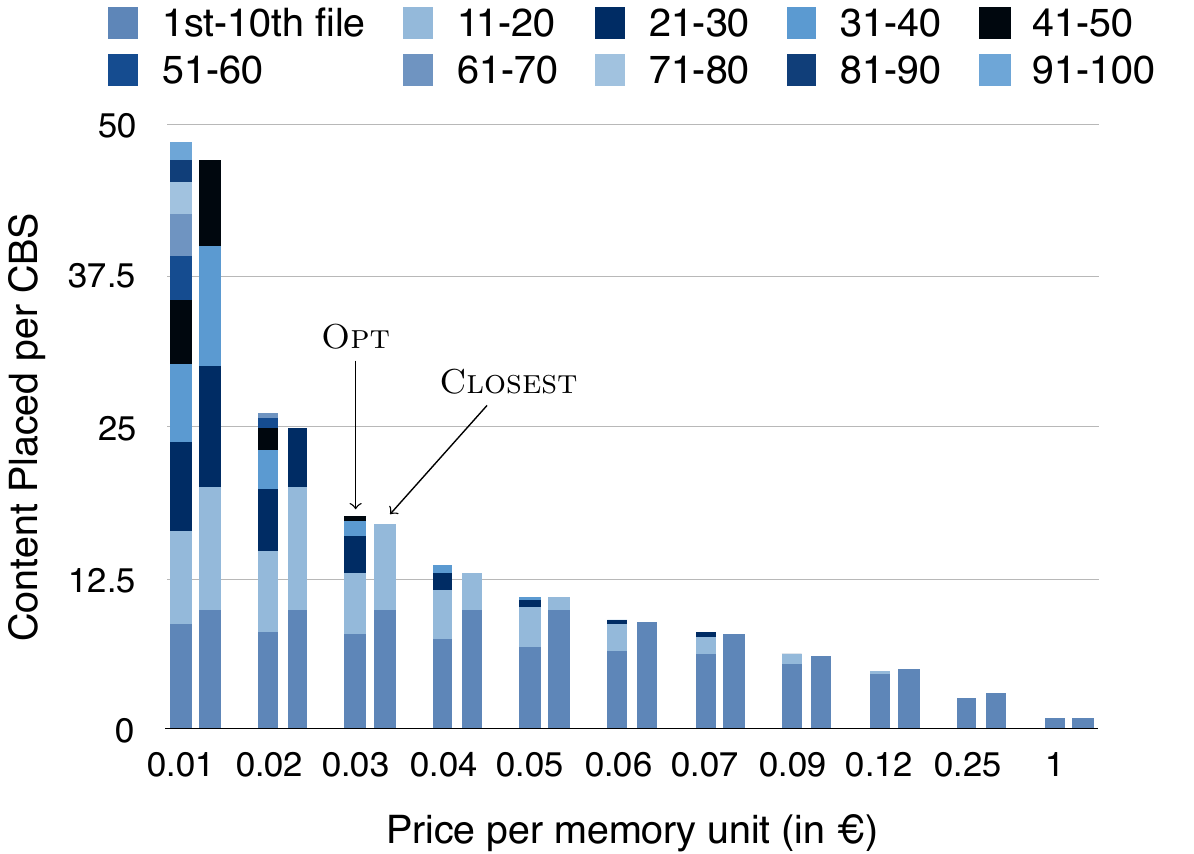}%
\label{fig:log-both-CP-price}%
\end{subfigure}\hfill%
\begin{subfigure}{.66\columnwidth}
\includegraphics[width=\columnwidth]{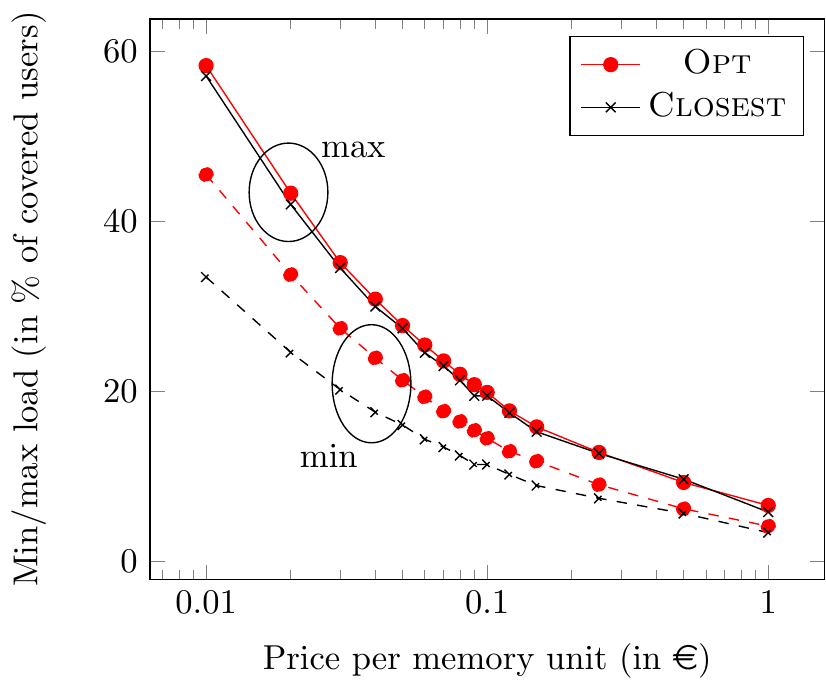}%
\label{fig:log-load-balance}%
\end{subfigure}%
\caption{Logarithmic savings function. (a): Hit ratio  in relation to price for different coverage radii for cases ii.a) and ii.b). (b): Cache lease and placement of popular files depending on cache unit price. Each left column represents case i.a), each right column is case i.b). (c) Cache lease and placement of popular files depending on cache unit price. Each left column represents case i.a), each right column is case i.b).}
\label{fig:log}
\end{figure*}

Here, we present the experimental results of case ii) in which the  savings function is the sum of logarithms.
Figure~\ref{fig:log}(a) %\ref{fig:log-distr-hr-price-z06} 
shows the  hit ratio for varying cache unit price both for the \textsc{Opt} (different coverage radii) and the \Closest policies. Due to the specific choice of the logarithmic savings function of case ii) the \Closest association gives identical  cache leasing and content placement results for all coverage radii. For every  coverage radius and every cache unit price, the \textsc{Opt} policy achieves a higher hit ratio than the \Closest policy. Furthermore, the higher the coverage radius in case ii.a), the higher the hit ratio. The hit ratio improvement can reach over $15$ percentage points using \textsc{Opt}. With increasing prices, caching becomes less profitable and the hit ratio decreases.

The  advantage to the hit ratio of the \textsc{Opt} policy (ii.a)  can be explained by the optimal content placement  shown in Figure~\ref{fig:log}(b). %\ref{fig:log-both-CP-price}. 
Each pair of columns represents cache leasing and content placement for a certain unit price. Each left column represents the decisions according to \textsc{Opt}, each right column the decisions according to \Closest. The height is the average amount of cache units leased per CBS. The inner sections of the columns represent the content placement in all of the CBSs: The lowest section are the 10 most popular files, the second lowest the files ranked 11 to 20 and so on. It can be seen that particularly for low cache unit prices, the diversity of cached content is higher in case ii.a) than in case ii.b). For the lowest price (0.01\euro{}), \Closest provides only content from the more popular half of the catalog, while \Opt places content from the tail of the catalog as well.

The main purpose of choosing the specific savings function in ii.a) is, however, the balancing of traffic load among the CBSs in order to avoid excess of resources by user overflow which will lead to service dissatisfaction. Figure~\ref{fig:log}(c) %\ref{fig:log-load-balance} 
shows that  the additional load (from the increase in hit ratio using \textsc{Opt}, see Figure~\ref{fig:log}(a)) %\ref{fig:log-distr-hr-price-z06}),
 is distributed to the less loaded CBSs. The two upper (solid) lines in the graph represent the maximum load of a CBS in relation to the overall covered population per CBS  both in  the cases ii.a) and ii.b). The two lower (dashed) lines are the minium loaded CBS. The maximum loaded CBSs in both ii.a) and ii.b) coincide  as the figure shows. The minimum loaded CBS of ii.a) is higher than the ii.b), showing that  excess users coming from the higher hit ratio %(Figure~\ref{fig:log}(a)) %\ref{fig:log-distr-hr-price-z06}) 
 are associated to the less loaded stations. %This follows from the higher load in these cases . 

The three plots show that  the \textsc{Opt} policy achieves an increase in hit ratio (good for both the CP and the MNO) while at the same time diversifying the cached content (good for the user) and avoiding an overload of CBSs (good for everybody).

\section{Conclusions}
\label{section:conclusion}

In this work, we propose a business model in which an MNO leases edge caches  to a CP. The  CP's objective is  to maximize its savings through offloading of traffic from its data centers to the wireless caches while limiting the cache leasing costs. The optimality of the  CP  decisions  depends on the MNO's user association policy as well as its pricing strategy.

 The problem is modelled  as a NLMIP taking the perspective of the CP. Network topology, association policy, pricing as well as CP savings function are general. Radio conditions and wireless node resource constraints can be included as soft constraints. An optimal solution to the general problem is found by applying  Benders decomposition. The solution method converges to the global optimum and allows for each of the players to take separate actions iteratively.

Extensive experiments for random network topologies  allow to compare the optimal CP decisions for different MNO association policies, cache prices, as well as  CP savings functions. In all versions of the problem, we have identified a unique price that maximizes the MNO revenue. It depends on how much the CP valorizes traffic offloading achieved by the edge caches. This information is included in the CP’s choice of the savings function. Another main conclusion is that  MNO association policies that adhere to CP actions and exploit multi-coverage opportunities achieve  higher  offloading benefits for a given monetary investment. All these results suggest that the CP and MNO can jointly develop cooperative
business models related to caching, that  lead to considerable 
economic as well as operational benefits for both parties.

\bibliographystyle{unsrt}
\bibliography{DocBib}

\begin{thebibliography}{10}

\bibitem{Cisco2016}
Cisco visual networking index: Global mobile data traffic forecast update,
  2015--2020 white paper.
\newblock White Paper, 2 2016.

\bibitem{Caire2013}
K.~Shanmugam, N.~Golrezaei, A.G. Dimakis, A.F. Molisch, and G.~Caire.
\newblock Femtocaching: Wireless content delivery through distributed caching
  helpers.
\newblock {\em Information Theory, IEEE Transactions on}, 59(12):8402--8413,
  Dec 2013.

\bibitem{Blaszczyszyn2014}
B.~B{\l}aszczyszyn and A.~Giovanidis.
\newblock Optimal geographic caching in cellular networks.
\newblock In {\em Communications (ICC), 2015 IEEE International Conference on},
  pages 3358--3363. IEEE, 2015.

\bibitem{Bastug2015}
E.~Ba{\c{s}}tu{\u{g}}, M.~Bennis, and M.~Debbah.
\newblock Cache-enabled small cell networks: Modeling and tradeoffs.
\newblock In {\em ISWCS 2014}, pages 649--653, Aug 2014.

\bibitem{Poularakis2016}
K.~Poularakis, G.~Iosifidis, I.~Pefkianakis, L.~Tassiulas, and M.~May.
\newblock Mobile data offloading through caching in residential 802.11 wireless
  networks.
\newblock {\em IEEE Transactions on Network and Service Management},
  13(1):71--84, 2016.

\bibitem{Douros2017}
V.~G. Douros, S.~E. Elayoubi, E.~Altman, and Y.~Hayel.
\newblock Caching games between content providers and internet service
  providers.
\newblock {\em Performance Evaluation}, 2017.

\bibitem{Araldo2016}
A.~Araldo, G.~Dan, and D.~Rossi.
\newblock Stochastic dynamic cache partitioning for encrypted content delivery.
\newblock In {\em Internet Teletraffic Congress (ITC) 2016}, 2016.

\bibitem{Poularakis2014b}
K.~Poularakis, G.~Iosifidis, and L.~Tassiulas.
\newblock Approximation algorithms for mobile data caching in small cell
  networks.
\newblock {\em IEEE Transactions on Communications}, 62(10):3665--3677, 2014.

\bibitem{Roberts2013}
J.~Roberts and N.~Sbihi.
\newblock Exploring the memory-bandwidth tradeoff in an information-centric
  network.
\newblock In {\em Proceedings of the 2013 25th International Teletraffic
  Congress (ITC)}, pages 1--9, Sept 2013.

\bibitem{Kelly1997}
F.~Kelly.
\newblock Charging and rate control for elastic traffic.
\newblock {\em Transactions on Emerging Telecommunications Technologies},
  8(1):33--37, 1997.

\bibitem{Dehghan2015}
M.~Dehghan, A.~Seetharam, Bo~Jiang, Ting He, Th. Salonidis, J.~Kurose,
  D.~Towsley, and R.~Sitaraman.
\newblock On the complexity of optimal routing and content caching in
  heterogeneous networks.
\newblock In {\em INFOCOM}, 2015.

\bibitem{Naveen2015}
K.P. Naveen, L.~Massoulie, E.~Baccelli, A.~Carneiro~Viana, and D.~Towsley.
\newblock On the interaction between content caching and request assignment in
  cellular cache networks.
\newblock In {\em ACM}, AllThingsCellular '15, pages 37--42. ACM, 2015.

\bibitem{Krolikowski2017}
J.\ Krolikowski, A.~Giovanidis, and M.~Di Renzo.
\newblock Fair distributed user-traffic association in cache equipped cellular
  networks.
\newblock In {\em WiOpt-CCDWN}, pages 1--6, 2017.

\bibitem{Bertsimas2005}
D.\ Bertsimas and R.\ Weismantel.
\newblock {\em Optimization over integers}.
\newblock Athena Scientific, 2005.

\bibitem{Paschos2016}
G.\ Paschos, E.\ Ba{\c{s}}tu\u{g}, I.\ Land, G.\ Caire, and M.~Debbah.
\newblock Wireless caching: Technical misconceptions and business barriers.
\newblock {\em IEEE Communications Magazine}, 54(8):16--22, 2016.

\bibitem{Bektacs2008}
T.\ Bekta{\c{s}}, J.\ Cordeau, E.\ Erkut, and G.~Laporte.
\newblock Exact algorithms for the joint object placement and request routing
  problem in content distribution networks.
\newblock {\em Computers \& Operations Research}, 35(12):3860--3884, 2008.

\bibitem{Schrijver1986}
A.~Schrijver.
\newblock {\em Theory of Linear and Integer Programming}.
\newblock John Wiley \& Sons, Inc., New York, NY, USA, 1986.

\bibitem{Geoffrion72}
A.~M. Geoffrion.
\newblock Generalized {B}enders decomposition.
\newblock {\em Journal of Optimization Theory and Applications}, 10:237--260,
  1972.

\bibitem{ElwalidMW06}
A.~Elwalid, D.~Mitra, and Q.~Wang.
\newblock Cooperative data-optical internetworking: Distributed multi-layer
  optimization.
\newblock In {\em INFOCOM}, 2006.

\bibitem{Mo2000}
J.~Mo and J.~Walrand.
\newblock Fair end-to-end window-based congestion control.
\newblock {\em IEEE/ACM Trans. Netw.}, 8(5):556--567, October 2000.

\bibitem{Neglia2017}
G.~Neglia, D.~Carra, and P.~Michiardi.
\newblock Cache policies for linear utility maximization.
\newblock In {\em {INFOCOM} 2017, May 1-4, 2017}, pages 1--9, 2017.

\bibitem{Liu2017b}
J.~Liu, B.~Bai, J.~Zhang, and K.~B. Letaief.
\newblock Cache placement in fog-rans: From centralized to distributed
  algorithms.
\newblock {\em IEEE Transactions on Wireless Communications},
  16(11):7039--7051, Nov 2017.

\end{thebibliography}
 
\end{document}